\documentclass[cits]{PoS}

\title{The Square Kilometre Array}

\ShortTitle{The Square Kilometre Array}

\author{\speaker{T.~Joseph~W.~Lazio}\thanks{Basic research in radio
	astronomy at the NRL is supported by 6.1 Base funding.}\\
        Naval Research Laboratory and Square Kilometre Array Program
Development Office, 4555 Overlook Ave.~\hbox{SW}, Washington, DC
20375-5351  USA\\
        E-mail: \email{lazio@skatelescope.org}}

\abstract{%
The Square Kilometre Array (SKA) is intended as the next-generation
radio telescope and will address fundamental questions in
astrophysics, physics, and astrobiology.  The international science
community has developed a set of Key Science Programs: 
(1)~Emerging from the Dark Ages and the Epoch of Reionization,
(2)~Galaxy Evolution, Cosmology, and Dark Energy,
(3)~The Origin and Evolution of Cosmic Magnetism, 
(4)~Strong Field Tests of Gravity Using Pulsars and Black Holes, and
(5)~The Cradle of Life/Astrobiology.
In addition, there is a design philosophy of ``exploration of the
unknown,'' in which the objective is to keep the design as flexible as
possible to allow for future discoveries.  Both a significant
challenge and opportunity for the \hbox{SKA} is to obtain a
significantly wider field of view than has been obtained with radio
telescopes traditionally.  Given the breadth of coverage of cosmic
magnetism and galaxy evolution in this conference, I highlight some of
the opportunities that an expanded field of view will present
for other Key Science Programs.}

\FullConference{Panoramic Radio Astronomy: Wide-field 1-2 GHz research on galaxy evolution - PRA2009\\
		 June 02 - 05 2009\\
		 Groningen, the Netherlands}

\begin{document}

\section{Introduction}\label{sec:jl.intro}

In the $20^{\mathrm{th}}$ Century, we discovered our place in the Universe.  We learned that it was much bigger than we imagined and much more exotic.  Beyond our Milky Way, the Universe is filled with galaxies---each their own island universe.  They range in size from dwarf galaxies barely able to survive near their larger neighbors to giant elliptical galaxies, orders of magnitudes larger than the Milky Way.  These galaxies of stars also contain a multitude of other components including gas with a wide range of temperatures; compact objects including white dwarfs, neutron stars, and black holes; and planets.  Over the course of the century, black holes moved from a theoretical curiosity to a well-recognized endpoint of stellar evolution and a likely fundamental component of the centers of galaxies, with the potential to power immense jets of relativistic particles that affect their surroundings.  By the end of the century, we were beginning to unveil the basic structure and processes of the Universe in which these objects are embedded, including evidence of its origin and the still mysterious properties of dark matter and dark energy.

Our probes of the Universe have expanded dramatically as well.
Electromagnetic radiation has been detected from celestial objects at
frequencies below~1~MHz ($\lambda \sim 300$~m) to energies exceeding
1~\hbox{TeV}.  Moreover, the range of possible signals has expanded
beyond just electromagnetic radiation.  Cosmic rays rain down on the
Earth, some with energies approaching those of macroscopic objects.
Gravitational radiation has been detected indirectly, and numerous
potential classes of sources have been suggested, with the expectation
that the Earth is awash in gravitational waves.  Neutrinos have been
detected from both the Sun and supernova \hbox{1987A}, and many of the
processes that generate high energy cosmic rays should also produce a
spectrum of high-energy neutrinos.

In the $21^{\mathrm{st}}$ Century, we seek to understand the Universe we inhabit.  To do so will require a suite of powerful new instruments, on the ground and in space, operating across the entire electromagnetic spectrum and for multiple decades.  Observations at centimeter- to meter wavelengths have provided deep insight to a wide range of phenomena ranging from the solar system to the most distant observable celestial emission. This long and rich record of important discoveries in the radio spectrum, including 3 Nobel prizes, has been possible since many of the relevant physical phenomena can only be observed, or understood best, at these wavelengths.  These phenomena include the cosmic microwave background (CMB), quasars, pulsars, gravitational waves, astrophysical masers, magnetism from planets through galaxies, the ubiquitous jets from black holes and other objects, and the spatial distribution of hydrogen gas, the predominant baryonic constituent of the Universe.  Moreover, through the invention of aperture synthesis, also recognized by the Nobel committee, radio astronomy has reached unprecedented levels of imaging resolution and astrometric precision, providing the fuel for further discovery.

With only a handful of exceptions, radio telescopes and arrays have been limited to apertures of about~$10^4$~m${}^2$, constraining, for instance, studies of the 21-centimeter hydrogen emission to the nearby Universe ($z \sim 0.2$)~\cite{cr04}.  Contemporaneous with the astronomical discoveries in the latter half of the $20^{\mathrm{th}}$ Century have been technological developments that offer a path to substantial improvements in future radio astronomical measurements.
Among the improvements are mass production of
centimeter-wavelength antennas enabling apertures potentially 100
times larger than previously available, fiber optics for the
transmission of large volumes of data, high-speed digital signal
processing hardware for the acquisition and analysis of the signals, and computational improvements leading to massive processing and storage.  These new technologies, combined with dramatically improved survey speeds and the other advances, can open up an enormous expanded volume of discovery space, providing access to many new celestial phenomena and structures, including 3-dimensional mapping of the web of hydrogen gas through much of cosmic history ($z \sim 2$).

\begin{figure}
\begin{center}
\includegraphics[width=0.9\textwidth]{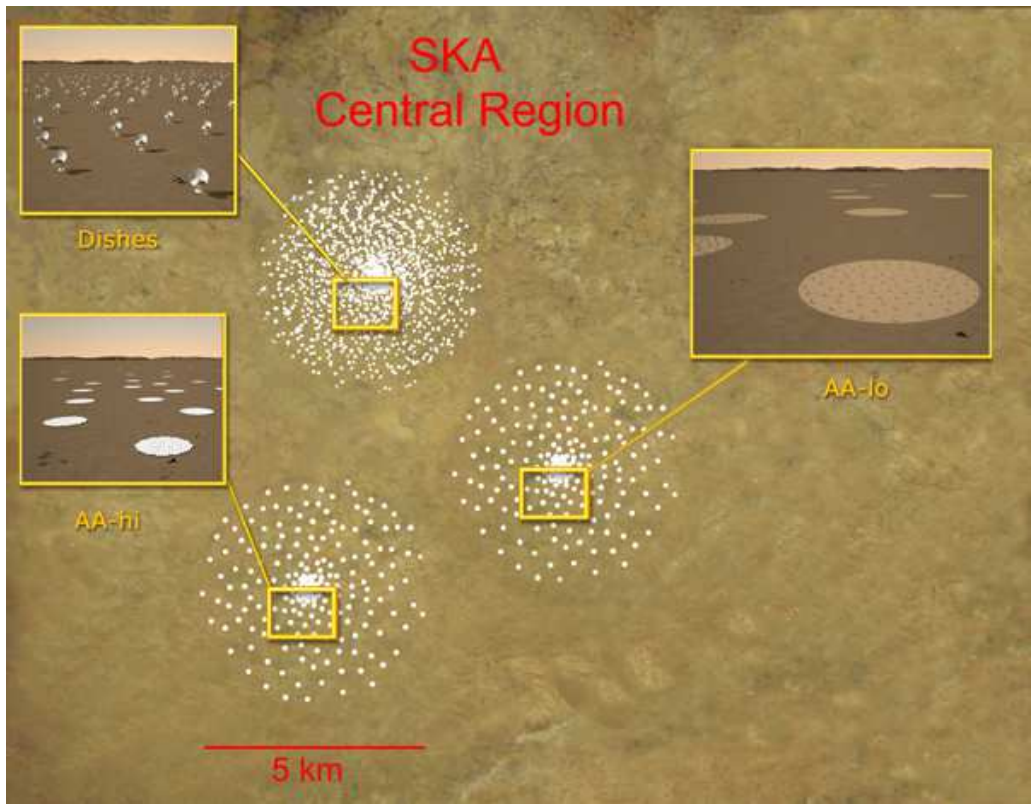}
\end{center}
\vspace*{-3ex}
\caption{An artist's impression of the core of the SKA illustrating
the various technologies over the frequency range 70~MHz to~10~GHz.
All of these technologies would enable various levels of wide-field
imaging.}
\label{fig:ska}
\end{figure}

The realization that radio astronomy was on the doorstep of a
revolutionary age of scientific breakthrough has led the international
community to investigate this opportunity in great detail over the
last decade.  That coordinated effort, involving a significant
fraction of the world's radio astronomers and engineers, has resulted
in the Square Kilometre Array (SKA) Program (Figure~\ref{fig:ska}), an international roadmap
for the future of radio astronomy over the next two decades and one
for which access to a wide field of view is an integral part of the science.

From its inception, development of the SKA Program has been a global
endeavor.  In the early 1990s, there were multiple, independent
suggestions for a ``large hydrogen telescope.''  It was recognized
that probing the fundamental baryonic component of the Universe much
beyond the local Universe would require a substantial increase in
collecting area.  The IAU established a working group in~1993 to begin
a worldwide study of the next generation radio observatory.  Since
that time, the effort has grown to comprise 19 countries and more than
50 institutes, including about~200 scientists and engineers.

%Significant milestones in the SKA Program have included the
%establishment of an international program office, selection of
%scientific specifications and a reference design, and identification
%of a short-list of suitable sites.  The SKA Program Development Office
%(SPDO) coordinates the on-going development work.  As many as 7
%different technical concepts for the SKA have been narrowed to a
%reference design, for which field-of-view expansion that would enable
%wide-field astronomical observations is a key component.  Finally, after a worldwide effort, two sites---one in the Karoo region of central South Africa and one in the state of Western Australia---have been identified as suitable for the site of the \hbox{SKA}.

\section{Key SKA Science}\label{sec:jl.science}

Over the past several years, there has been extensive activity related
to developing a detailed science case for the \hbox{SKA}, culminating
in the SKA Science Book \cite{cr04}.  Highlighting the SKA Science
Case are Key Science Projects (KSPs), which represent unanswered
questions in fundamental physics, astrophysics, and astrobiology.
Furthermore, each of these projects has been selected using the
criterion that it represents science that is either unique to the
\hbox{SKA} or in which the SKA will provide essential data for a
multi-wavelength analysis \cite{g04c}.
The KSPs are 

\begin{description}
\item[Emerging from the Dark Ages and the Epoch of Reionization]%
The ionizing ultra-violet radiation from the first stars and galaxies
produced a fundamental change in the surrounding intergalactic medium,
from a nearly completely neutral state to the nearly completely
ionized Universe in which we live today.  The most direct probe of
this Epoch of Re-ionization (EoR), and of the first large-scale
structure formation, will be obtained by imaging neutral hydrogen and
tracking the transition of the intergalactic medium from a neutral to
ionized state.  Moreover, as the first galaxies and AGN form, the SKA
will provide an unobscured view of their gas content and dynamics via
observations of highly redshifted, low-order molecular transitions
(e.g., CO).
  
\item[Galaxy Evolution, Cosmology, and Dark Energy]%
Hydrogen is the fundamental baryonic component of the Universe.  The
SKA will have sufficient sensitivity to the 21-cm hyperfine transition
of {H}\,\textsc{i} to detect galaxies to redshifts $z > 1$.  One of the key
questions for $21^{\mathrm{st}}$ Century astronomy is the assembly of
galaxies; the SKA will probe how galaxies convert their gas to stars
over a significant fraction of cosmic time and how the environment
affects galactic properties.  Simultaneously, baryon acoustic
oscillations (BAOs), remnants of early density fluctuations in the
Universe, serve as a tracer of the early expansion of the
Universe. The SKA will assemble a large enough sample of galaxies to
measure BAOs as a function of redshift to constrain the equation of
state of dark energy.

\item[The Origin and Evolution of Cosmic Magnetism]%
Magnetic fields likely play an important role throughout astrophysics,
including in particle acceleration, cosmic ray propagation, and star
formation.  Unlike gravity, which has been present since the earliest
times in the Universe, magnetic fields may have been generated
essentially \textit{ab initio} in galaxies and clusters of galaxies.  By
measuring the Faraday rotation toward large numbers of background
sources, the SKA will track the evolution of magnetic fields in
galaxies and clusters of galaxies over a large fraction of cosmic
time.  The SKA observations also will seek to address whether magnetic
fields are primordial and dating from the earliest times in the
Universe or generated much later by dynamo activity.
  
\item[Strong Field Tests of Gravity Using Pulsars and Black Holes]%
With magnetic field strengths as large as $10^{14}$~\hbox{G}, rotation
rates approaching 1000~Hz, central densities exceeding
$10^{14}$~g~cm${}^{-3}$, and normalized gravitational strengths of
order 0.4, neutron stars represent extreme laboratories.  Their
utility as fundamental laboratories has already been demonstrated
through results from observations of a number of objects. The SKA will
find many new millisecond pulsars and engage in high precision timing
of them in order to construct a Pulsar Timing Array for the detection
of nanohertz gravitational waves, probing the space-time environment
around black holes via both ultra- relativistic binaries (e.g.,
pulsar-black hole binaries) and pulsars orbiting the central
supermassive black hole in the centre of the Milky Way, and probe the
equation of state of nuclear matter.
  
\item[The Cradle of Life]%
The existence of life elsewhere in the Universe has been a topic of
speculation for millennia.  In the latter half of the
$20^{\mathrm{th}}$ Century, these speculations began to be informed by
observational data, including organic molecules in interstellar space,
and proto-planetary disks and planets themselves orbiting nearby
stars. With its sensitivity and resolution, the SKA will be able to
observe the centimeter-wavelength thermal radiation from dust in the
inner regions of nearby proto-planetary disks and monitor changes as
planets form, thereby probing a key regime in the planetary formation
process.  On larger scales in molecular clouds, the SKA will search
for complex prebiotic molecules.  Finally, detection of transmissions
from another civilization would provide immediate and direct evidence
of life elsewhere in the Universe, and the SKA will provide sufficient
sensitivity to enable, for the first time, searches for unintentional
emissions or ``leakage.''
\end{description}  
 
In addition to the KSPs listed, and recognizing the long history of
discovery at radio wavelengths (pulsars, cosmic microwave background,
quasars, masers, the first extrasolar planets, etc.), the
international science community also recommended that the design and
development of the SKA have ``Exploration of the Unknown'' as a
philosophy.  Wherever possible, the design of the telescope is being
developed in a manner to allow maximum flexibility and evolution of
its capabilities to probe new parameter space (e.g., time-variable
phenomena that current telescopes are not well-equipped to detect).
This philosophy is essential as many of the outstanding questions of
the 2020--2050 era---when the SKA will be in its most productive
years---are likely not even known today.

\section{Opportunities for Panoramic SKA Science}\label{sec:jl.pra}

Many of the papers in this volume illustrate far better than I could
the opportunities for Panoramic SKA Science, particularly in the areas
of cosmic magnetism and galaxy structure and evolution via
H\,\textsc{i} observations.  Consequently, and similar to my approach
in the conference itself, I shall focus on opportunities for
wide-field observations as they concern some of the other SKA KSPs.

\subsection{Emerging from the Dark Ages and the Epoch of
	Reionization}\label{sec:jl.eor}

The primary focus of this KSP is tracking the transition from the
Universe's largely neutral state to its currently nearly completed
ionized state.  Wide-field observations will be both important and
natural as the key observations will be of the highly-redshifted
H\,\textsc{i} line at frequencies below~200~MHz, which will be carried
out using dipole-based sparse aperture arrays.  Dipoles have fields of
view that can exceed $\pi$~sr easily, but the relevant frequencies are
below the nominal focus of this conference.

A potential and important secondary observation that could be
conducted near~1~GHz, however, would be of the synchrotron radiation
from the first galaxies \cite{m09}.  Copious numbers of massive stars
would have likely formed within these first galaxies and then exploded
soon thereafter as supernovae.  If the interstellar magnetic fields of
these galaxies have developed sufficiently, the galaxies will emit
synchrotron radiation as a result of cosmic rays accelerated by the
supernova remnants from these first massive stars.  While it is not
yet known if these galaxies will be detectable, the radio-far infrared
correlation for star-forming galaxies is now known to hold at least
out to a redshift $z \approx 3$ \cite{shd+09}.  If it continues to
hold to $z \approx 6$, then radio observations would be a powerful
means of probing dust-enshrouded first galaxies and wide-field
observations would naturally allow for large volumes of the Universe
to be sampled quickly.

\subsection{Fundamental Physics Using Observations of Pulsars and
	Black Holes}\label{sec:jl.psr}

Wide-field capabilities that enable the SKA to access a substantial
solid angle will be important for pulsar studies, even though pulsars
are point sources so that ``panoramic imaging'' per se of them is
unlikely to be profitable.

Fundamental physics constraints are derived from pulsar observations
via long-term timing programs that measure precisely the times of
arrival of the pulses.  A significant constraint on the utility of
radio pulsars is the scarcity of ``useful'' pulsars.  For instance,
until recently, the most significant constraints on the nuclear
equation of state derived from radio pulsars resulted from the first
millisecond pulsar, PSR~B1937$+$21, discovered in the \emph{early
1980s}.  Similarly, many of the tests of theories of gravity and for
gravitational wave emission rely on one or a few objects.  Recent
surveys have begun to demonstrate the potential for vastly increasing
the number of radio pulsars and thereby increasing the number of
``useful'' systems.

Perhaps the best example of the impact of increasing field of view for
pulsar surveys is the Parkes Multibeam Survey \cite{mlc+01}.
By installing a
multiple feed horn system on the Parkes antenna, the effective field
of view was increased by a factor of~13.  The resulting survey
essentially doubled the total number of pulsars.  (See also
\S\ref{sec:jl.dynamic}.)  Future field-of-view expansion technologies
(e.g., phased array feeds or dense aperture arrays) coupled with the
vastly increased sensitivity of the SKA offer promise for an even
larger yield.

The impact of a wide field of view for pulsar timing and monitoring
programs is less clear.  In principle, a telescope with a sufficiently
wide field of view could time multiple pulsars simultaneously,
yielding an improved ``throughput.''  In practice, the current
estimates of the density on the sky of ``useful'' pulsars is
sufficiently low that only dense aperture arrays are likely to have a
field of view that could be large enough to time multiple pulsars
simultaneously, except perhaps in special regions of the sky.
Moreover, in order to mitigate interstellar propagation effects,
timing observations have to be carried out over a relatively large
frequency range (e.g., 0.8--3~GHz), wider than what dense aperture
arrays are currently thought to be able to achieve.

\subsection{Cradle of Life/Astrobiology}\label{sec:jl.col}

One of the key assumptions in the search for life elsewhere in the
Universe, particularly in searches for life within the solar system,
is that other life is likely to be based on carbon chemistry (i.e.,
``organic''), like life on Earth is.  Prime support for this approach
is that the vast majority of multi-atom molecules in interstellar
space contain carbon, including a number of complex organic species
\cite[and references within]{shjlr06,bgmmcs08,bgmmcs09}.  As the
number of atoms increases, the rotational and vibrational transitions
tend to shift to lower frequencies, and searches for and studies of
complex organic molecules have relied upon observations below~2~GHz.
Consequently, a wide field-of-view at frequencies around~1~GHz could
be quite valuable for conducting surveys of molecular clouds for
complex organic molecules; such observations would find a natural
complement in ALMA observations.

%ref
Direct evidence for life elsewhere in the Universe would be the
detection of signals from another technological civilization.  Two
examples from our own civilization are cell phone transmissions and
aeronautical navigation, both of which make use of frequencies
around~1~GHz, though neither are strong enough to be detectable over
interstellar distances (even with the SKA!).  More generally, the
``waterhole'' between~1.4 and~1.7~GHz has been a focus of numerous
previous searches for extratestrial transmissions, as it has been
argued that any technological civilization capable of trying to
communicate over interstellar distances would certainly know about the
H\,\textsc{i} line at~1.4~GHz and the OH lines around~1.7~GHz.

One approach for searching for extraterrestial intelligence (SETI) is
to monitor a ``habstar,'' a star that might be orbited by a terrestrial
planet(s) within the star's habitable zone \cite{tt03}.  Much like pulsar timing,
being able to monitor multiple habstars would increase the throughput
of SETI observations; the key contrast between pulsar and habstar
observations is that the density on the sky of suitable main sequence
stars is sufficiently high that most, if not all, fields of view will
include more than one habstar.

\subsection{The Dynamic Radio Sky}\label{sec:jl.dynamic}

A series of discoveries over the past decade have both illustrated and
emphasized that the time domain has been explored only poorly at radio
wavelengths \cite{bsb+07,hbl+07,hlkrmy-z05,mll+06}.  Although time
resolutions approaching 1~ns have been achieved \cite{hkwe03},
typically these have been obtained only on relatively narrow fields of
view.  The challenge and opportunity for the \hbox{SKA}, and
consistent with the ``exploration of the unknown'' design philosophy,
is to obtain both high time resolution and access to a significant
solid angle.

Some of these observations might naturally happen in the course of
pulsar surveys (\S\ref{sec:jl.psr}); indeed, the discovery of rotating
radio transients, a new class of radio-emitting neutron stars,
resulted from the novel processing of a pulsar survey \cite{mll+06}.
Other types of transient surveys and exploration programs might
utilize wide fields of view in different manners, however.

We provide two examples to illustrate the potential range of
applications of a wide field of view:
\begin{enumerate}
\item Extreme scattering events (ESEs) are a class of dramatic flux
density variations ($\sim 50$\%) of extragalactic sources caused by
intervening plasma lenses \cite{fdjh87}.  The initial surveys for ESEs
observed only a relatively small number of the strongest, most compact
sources on the sky.  Yet within even a modest field of view, if the
full field of view can be imaged, are potentially tens to hundreds of
sources.  A potential ESE search program could be conducted by
surveying a significant solid angle with a regular cadence and
constructing light curves of all of the sources within the survey
region.  Clearly an expanded field of view would determine the total
number of sources that could be monitored.

\item Many low-mass stars (spectral types K and~M) show significant
``radio activity,'' often with radio flares or bursts on short time
scales \cite{ob08}.  This radio emission is thought to be linked to
coronal processes on the stars, likely closely coupled to the magnetic
field structure.  Study of the coronal processes in these extreme
cases may provide understanding of solar processes, which could impact
not only astrophysics but aspects of the Earth-Sun connection as
well.  Most, if not all, of the strongly ``radio active'' stars in the
solar neighborhood are known, but the typical separation on the sky is
fairly large.  Similar to the case for pulsar timing, monitoring a
large number of low-mass stars for radio bursts would have a much
higher throughput if access to a wide field of view becomes possible.
\end{enumerate}
We emphasize that these are only two possible examples, chosen to
illustrate the possible range of transient survey programs.  The
actual impact of the field of view on any transient program will also
depend upon the temporal characteristics of the transients being
targeted, their luminosity function, and distribution on the sky, to
the extent that these parameters are known.

\end{document}